\documentclass[referee]{raa}           
\usepackage{graphicx,times}
\usepackage{natbib}
\usepackage{amssymb,amsmath}
\bibpunct{(}{)}{;}{a}{}{,}

\usepackage[pagebackref=true]{hyperref}
\hypersetup{colorlinks = true, linkcolor = green, anchorcolor = red, citecolor = blue, filecolor = red, pagecolor = red, urlcolor = red}

\begin{document}

   \title{Photometric analysis of two extreme low mass ratio contact binary systems}

 \volnopage{ {\bf 20XX} Vol.\ {\bf X} No. {\bf XX}, 000--000}
   \setcounter{page}{1}

   \author{Surjit S Wadhwa\inst{1}, Nick F H Tothill\inst{1}, Ain Y DeHorta\inst{1}, Miroslav Filipović \inst{1}
   }

   \institute{School of Science, Western Sydney University, Locked Bag 1797, Penrith, NSW 2751, Australia {\it  19899347@student.westernsydney.edu.au}\\
\vs \no
   {\small Received 20XX Month Day; accepted 20XX Month Day}
}

\abstract{Multi band photometry and light curve analysis for two newly recognized contact binary systems, TYC 6995-813-1 and NSVS 13602901 are presented. Both were found to be of extreme low mass ratios 0.11 and 0.17, respectively. The secondary components of both systems show evidence of considerable evolution with elevated densities as well as both luminosity and radii well above their main sequence counterparts. Even in the absence of significant spot activity at least one of the systems, TYC 6995-813-1, shows features of magnetic and chromospheric activity. TYC 6995-813-1 is also determined to be a potential merger candidate with its current separation near the theoretical instability separation. 
}

   \authorrunning{Wadhwa et al. }            
   \titlerunning{Extreme low mass ratio contact binaries}  
   \maketitle

%
\section{Introduction}           
\label{sect:intro}

The merger of contact binary star components are thought to result in a transient event referred to as a red nova. Estimates suggest that up to 1 in 500 stars in the Galaxy maybe contact binaries \citep{2007MNRAS.382..393R}. The merger event is relatively rare and until recently there had been no confirmed observed events.  The initial eruption of V1309 Scorpii (V1309 Sco) \citep{2008IAUC.8972....1N} was thought to be a classical nova eruption because of the rapid increase in brightness.  In the days following the initial eruption, V1309 Sco did not behave as expected, rapidly cooling in a matter of days \citep{2010A&A...516A.108M}. The spectrum became redder (cooler), as opposed to hotter as in a classical nova, suggesting the event was most likely a luminous red nova. Although not known prior to the event, the progenitor of V1309 Sco had been in the field of view of the OGLE (Optical Gravitational Lensing Experiment) for a number of years prior to the outburst. Photometric analysis of the OGLE data indicated that the progenitor was indeed a contact binary system and that V1309 Sco was indeed a merger event between the components of a contact binary system with a resultant rapidly rotating single star \citep{2011A&A...528A.114T}. Recently there has been a rise in the interest of contact binary star mergers with a number of new projects exploring the theoretical \citep{2021MNRAS.501..229W} and observational aspects \citep{2021MNRAS.502.2879G}

There has been heightened interest in orbital stability of contact binary systems as a way of identifying of potential merger candidates since the eruption of V-1309. The interplay of orbital and spin angular momentum is critical in maintaining orbital stability with instability likely if orbital angular momentum decreases below three times the total spin angular momentum of the components \citep{1980A&A....92..167H}. Determination of angular momentum is dependent on accurate determination of absolute parameters and contact binary systems provide one of the best avenues for determining the basic absolute parameters such as masses and radii of stars. Previous work \citep{2007MNRAS.377.1635A} has shown that in most instances, orbital instability is likely to occur in systems with extreme low mass ratios.

In this study we report the multi-band photometry of two recently identified contact binary systems. We performed simultaneous multi-band analysis of the light curve using a recent version of the \citet{1971ApJ...166..605W} code. We show that both systems are suitable for light curve analysis with complete eclipses, both are of extreme low mass ratio and both have significantly brighter and larger secondary components relative to their main sequence counterparts. In addition, where possible, we explore the evolution of the systems and other astrophysical indicators of potential instability.

\section{Photometry and Light Curves}
Images were acquired in B,V and R bands using the Western Sydney University (WSU) 0.6m telescope equipped with a SBIG 8300T CCD camera and standard filters. Images were plate solved using the American Association of Variable Star Observers (AAVSO) VPHOT engine and differential photometry was also performed using the VPHOT engine. The photometry report from the VPHOT engine gives an estimate of the error and all observations where the reported error exceeded 0.01 magnitude were excluded. We used the method of \citet{1956BAN....12..327K} to derive new times of minima and new period was derived using the PSeach utility available through the AAVSO. The photometric data was folded using the derived period and time of minima. The deeper of the two eclipses was designated as the primary eclipse at phase zero. The maximum brightness for each pass-band at phase 0.25 and 0.75 were determined using parabolic fit for observations from phase 0.24-0.26 and 0.74-0.76 respectively.  In order to determine the absolute magnitude of the primary (see below) the V band magnitude at mid secondary eclipse was taken as the average magnitude between phases 0.49-0.51 as all systems showed complete eclipse during this part of the phase cycle.  Parabolic fit between phase 0.95 and 0.05 combined with the maximum magnitude was used to determine the amplitude of each system. As both systems being imaged are relatively new discoveries no orbital period change analysis was possible given only scant (or nil) historical observations.
\subsection{TYC 6995-813-1}
TYC 6995-813-1 ($\alpha_{2000.0} = 00\ 06\ 49.98$, $\delta_{2000.0} = -35\ 37\ 29.1$) was identified as a contact binary both by the Catalina survey \citep{2017MNRAS.469.3688D} and the All Sky Automated Survey - Super Nova survey (ASAS-SN) \citep{2018MNRAS.477.3145J} with a preliminary period of 0.38318 days. Differential photometry was performed with TYC 6995-815-1 as the comparison star and 2MASS 00071379-3534495 as the check star. The amplitude in V band was approximately 0.34 magnitudes (12.08 - 12.42) while the difference between the primary and secondary eclipses, as expected, was small at 0.04 magnitudes. The red band amplitude was 0.32 magnitudes (11.97 - 12.29) while the blue band amplitude was 0.37 magnitudes (12.63 - 13.00). The light curve demonstrates a complete eclipse lasting approximately 28 minutes. A single new minimum time for the primary eclipse was determined as HJD 2458797.9902 (0.0002) and a more refined period of 0.383130 (0.00002) days calculated.
\subsection{NSVS 13602901}
NSVS 13602901 ($\alpha_{2000.0} = 16\ 46\ 09.83$, $\delta_{2000.0} = -03\ 52\ 14.3$) was recognised as a contact binary systems both by the All Sky Automated Survey \citep{1997AcA....47..467P} and Northern Sky Variable Survey \citep{2003AAS...203.5703W} with a preliminary period of 0.52379 days. As per TYC 6995-813-1 images were acquired using the UWS telescope and photometry performed using the VPHOT engine with TYC 5054-1601-1 as the comparison star and TYC 5058-642-1 as the check star. The V band amplitude was 0.4 magnitudes (11.83 - 12.23) while difference in the eclipse depths was again 0.04 magnitudes. The red band amplitude was 0.39 magnitudes (11.47 - 11.86) while the blue band amplitude was 0.41 magnitudes (12.45 - 12.86). The eclipses are complete lasting approximately 69 minutes. A single new time for the primary minimum was determined as HJD 2458737.89608 (0.00003) while the period was refined to 0.5238905 (0.00002) days.
\section{Light Curve Analysis}
Reliable light curve solutions of contact binary systems without spectroscopic mass ratios is only possible in cases where complete eclipses are present to constrain the the highly correlated geometric parameters such as the mass ratio ($q$), inclination ($i$) and fill-out ($f$) \citep{2005Ap&SS.296..221T}. Both our systems show complete eclipses and are thus suitable for photometric analysis without needing spectroscopic observations. All light curves were solved using a recent (2010) version of the Wilson-Devinney code (WD) \citep{1971ApJ...166..605W}. As there was no appreciable variation in the two brightest magnitudes (O’Connell Effect) only unspotted solutions were sought for both systems. As geometric elements account almost entirely for the shape of contact binary light curves \citep{1993PASP..105.1433R} certain parameters were fixed for both systems. As the components of contact binary systems are surrounded by a common we applied simple reflection treatment \citep{1969PoAst..17..163R} and fixed the bolometric albedos $A_1 = A_2 = 0.5$ and the gravity darkening coefficients $g_1 = g_2 = 0.32$ \citep{1967ZA.....65...89L}. Limb darkening coefficients for each pass-band were interpolated from \citet{1993AJ....106.2096V} and logarithmic law applied as per \citet{2015IBVS.6134....1N}. The other fixed parameter in all cases was the effective temperature of the primary ($T_1$). This was estimated firstly by estimating the absolute magnitude of the primary ($Mv_1$) by using the mid-eclipse magnitude ($m$) from the light curve (as this represents light from the primary only), distance ($D$) \citep{2018yCat.1345....0G}, and interstellar reddening [E(B-V)] \citep{1998ApJ...500..525S} as follows:
\begin{equation}
    M_v = m - (5*log_{10}(D/10)) + 3.1*E(B-V)
\end{equation}
Temperature of the primary was then determined from calibration for main sequence stars from \citet{2000asqu.book.....C}. There are various approaches to estimate the effective temperature of the primary with wide variations possible. In the case of NSVS 1360291 the \citet{2018yCat.1345....0G} reports the effective temperature as 5725K while the LAMOST catalog \citep{2019yCat.5164....0L} suggests likely temperature as approximately 7300K. As geometric parameters almost exclusively determine the shape of contact binary light curves variations in the estimate of $T_1$ will have no significant effect on the light curve solution with respect to the photometric mass ratio and other major parameters such as inclination and fill-out \citep{1993PASP..105.1433R}.

Geometric parameters such as the mass ratio, inclination and degree of contact are the dominant variables determining the shape of contact binary light curves with gravity/limb darkening and reflection playing a minimal role \citep{1993PASP..105.1433R}. As noted by \citet{2005Ap&SS.296..221T} significant correlation exists between the mass ratio and the other two major geometric parameters and reliable photometric solutions cannot be obtained, except circumstances of a total eclipses, when the mass ratio and the other two combinations are treated as free. The systemic mass ratio search grid method described by \citet{1982A&A...107..197R} involves obtaining the best solution for a range of fixed values of the mass ratio $q$, with other parameters such as the temperature of the secondary ($T_2$), potential (fill-out), and inclination acting as free parameters. The solution with the smallest total error between the observed and modelled light curve is then selected as representing the true mass ratio.
\citet{2005Ap&SS.296..221T} have previously shown that such a technique when used in cases of complete eclipses will yield the true mass ratio. For both systems simultaneous multi-band solutions were obtained for various fixed incremental (0.01) values of $q$ to initially find a approximate solution. Finer grid search at increments of 0.001 for the value of $q$ was conducted near the approximate value to obtain the true mass ratio. In the final iteration the mass ratio was also made a free parameter and the software reported standard deviations were recorded as the errors in the light curve solution. Fitted and observed light curves and mass ratio search grids for TYC 6995-813-1 and NSVS 13602901 are shown in figures 1 and 2 respectively while 3D representations are illustrated in figures 3. The light curve solution along with absolute and other astrophysical parameters (see below) are summarised in table 1.
\section{Absolute and Astrophysical Parameters}
\citet{2013MNRAS.430.2029Y}  showed that the primary components of contact binaries in terms of mass, radius and luminosity are much closer to that of the normal zero age main sequence (ZAMS) stars than the secondary components. Despite their low masses, the secondary component are significantly brighter and larger than to their ZAMS or terminal age main sequence (TAMS) counterparts. Using equation 1 it is possible to determine the absolute magnitude of the primary and using the mass luminosity relation of \citep{1991Ap&SS.181..313D} and the light curve solution it is possible to determine the masses of both components. The error in he mass of the primary was estimated from the error in the distance \citep{2018yCat.1345....0G} and subsequent errors estimated through error propagation. As the total luminosity of the system is known and the luminosity of the primary determined as described the luminosity of the secondary can be easily derived. The hallmark of a contact binary system is that both components overflow their Roche lobes and as such their relative radii are dependent on the Roche geometry and the mass ratio. As part of the light curve solution the WD code provides the fractional radii of the component stars in three orientations. In this study we use the geometric mean of the three directional radii as an estimate of the fractional radii of the primary ($r_1$) and secondary ($r_2$). From Kepler's third law, mass ratio and period it is possible to determine the separation ($A$) and from the separation the absolute radii of the components as per \citep{2005JKAS...38...43A} $R_{1,2}$ = $A*r_{1,2}$ can be calculated.

\citet{2013MNRAS.430.2029Y} compiled these basic parameters for 100 well studies contact binary systems and showed good agreement between the luminosities of the primary component and main sequence stars of similar mass. With respect to the radii they showed that the primaries have somewhat larger radii than ZAMS stars. Unlike the primaries the luminosities and radii of the secondary do not correlate with main sequence stars of similar mass. They found that all secondary components were considerably brighter and larger than TAMS stars. The relative radii and luminosities and their relationship to ZAMS and TAMS are summarised in table 1 and confirm the general reported findings.

\citet{2013MNRAS.430.2029Y} argue that given the distorted nature of the secondaries they must have different structure and evolutionary path relative to single stars. They argue that the interior of the secondaries are quite different from main sequence stars because their initial mass was significantly higher than the present mass as significant mass loss (or mass transfer) has occurred from the secondary to the primary. This situation means the present secondary component had a different evolutionary path compared to main sequence stars. Through detailed modelling they showed that the present secondary was in fact larger and the initial primary which lost mass to the initial smaller secondary through a semi-detached phase. The initial secondary grew larger while the initial primary grew smaller. Eventually, the initial secondary became larger than the mass losing primary and assumed the role of the present primary. Due to loss of mass from the outer layers of the current secondary, the internal structure of the current secondary is thought to be considerably denser than that of the primary. Modelling by \citet{2004A&A...414..317K} showed that the difference in the densities $\Delta\rho = \rho_1 - \rho_2$ ($\rho_{1,2}$ = densities of the primary and secondary respectively) is always negative for all contact binary systems. Knowing the period, radii, and mass ratio, the mean density of each component can be easily computed. The density (in $gcm^{-3}$) of the components can be expressed as a function of the mass ratio, period and radii of the components \citep{1981ApJ...245..650M} and the difference between the densities of the components can be expressed as follows:
\begin{equation}
    \Delta\rho = \frac {0.0189q}{R_2^3(1+q)P^2} - \frac {0.0189}{R_1^3(1+q)P^2}
\end{equation}
where $q$ is the current mass ratio, $R_{1,2}$ radii of the current primary and secondary and $P$ is the period in days. $\Delta\rho$ was calculated for each of the systems and reported in table 1 and as predicted it is strongly negative.

Contact binary systems are known to be magnetically active and hence potentially chromospherically active. Activity signals are common in contact binary systems. One of these is the asymmetry in the light curve maxima (O’Connell effect). Apart from star spot modelling, light curve analysis provides little indication of choromospheric activity although much clearer indicators such as certain emission lines are more specific for chromospheric activity. In the optical waveband the chromospheric emissions are obscured by the intense photospheric activity. The same however is not true for far ultraviolet wavelengths, especially in the case of late type dwarfs which make up most of the contact binary systems \citep{2010PASP..122.1303S}. The GALEX (Galaxy Evolution Explorer) satellite imaged the sky in two photometric band-passes: a near-ultraviolet band (NUV) with an effective wavelength of 2316{\AA} and a far band (FUV) centered on 1539{\AA} \citep{2007ApJS..173..682M}. Only the FUV band is useful for the detection of chromospheric activity as the NUV band remains contaminated by photospheric activity \citep{2010PASP..122.1303S}. A large volume of data on the strength of active chromospheric emission lines for a large number of late type dwarf stars exists as a result of the Mount Wilson HK project \citep{1978PASP...90..267V, 1991ApJS...76..383D}. \citet{2010PASP..122.1303S} used the  GALEX FUV magnitudes ($m_{FUV}$) combined with the HK survey data to investigate the sensitivity of FUV brightness as an indicator of chromospheric activity.

The most accepted measure of chromospheric emission strength is the fraction of the star's bolometric luminosity emitted in the active H and K lines \citep{1984ApJ...279..763N}. The value is normally expressed as $log R_{HK}^{'}$ and stars with $log R_{HK}^{'} \geq -4.75$ are considered more active \citep{1996AJ....111..439H}. Using the large database of $log R_{HK}^{'}$ values of dwarf stars \citet{2010PASP..122.1303S} matched the observations with GALEX magnitudes to derive two key relationships:
\begin{equation}
    (m_{FUV}-B)_{base} = 6.73(B-V) + 7.43
\end{equation}
which defines the $m_{FUV}-B$ for stars with the weakest emissions and low activity. This is the used as a correction term when defining the colour excess:
\begin{equation}
    \Delta(m_{FUV}-B) = (m_{FUV}-B) - (m_{FUV}-B)_{base}
\end{equation}
They showed that for active stars i.e those with $log R_{HK}^{'} \geq -4.75$ the colour excess was generally less than -0.5 and more frequently less than -1.0. Less active stars had colour excess values significantly higher than -0.5. Only one of our systems, TYC 6995-813-1, was observed by GALEX with the reported $m_{FUV}$ of 20.85. Using this and the relationships above (we use the published B-V value from ASAS-SN database) we derive the UV colour excess of -2.89 indicating underlying significant choromospheric activity even in the absence of significant spot activity.\\

\begin{table}[h!]
	\centering
	
	\label{tab:RRA T1}
	\begin{tabular}{|c|c|c|}
          \hline
          Parameter & TYC 6995-813-1&NSVS 13602901\\
          \hline
          Distance (pc)&$498.0\pm14$&$6114.4\pm40$\\
          $M_1/M_{\sun}$&$1.23\pm0.01$&$1.19\pm0.02$\\
          $M_2/M_{\sun}$&$0.135\pm0.01$&$0.203\pm0.01$\\
          $T_1$(K)(Fixed)&6300&6250\\
          $T_2$(K)&$6235\pm27$&$6222\pm18$\\
          Inclination($^{\circ}$)&$84.03\pm1.51$&$83.59\pm1.15$\\
          Fill-out($f$)\%&$72\pm2$&$44\pm2$\\
          Potential($\Omega$)&$1.941\pm0.002$&$2.11\pm0.007$\\
          $r_1$(mean)&0.633&0.554\\
          $r_2$(mean)&0.288&0.260\\
          Mass Ratio ($M_2/M_1$)&$0.111\pm0.002$&$0.171\pm0.002$\\
          $A$($R_{\sun}$)&$2.46\pm0.01$&$3.05\pm0.01$\\
          $R_1/R_{\sun}$&$1.46\pm0.01$&$1.69\pm0.01$\\
          $R_2/R_{\sun}$&$0.60\pm0.01$&$0.79\pm0.01$\\
          $L_1/L_{\sun}$&$2.29\pm0.04$&$2.05\pm0.04$\\
          $L_2/L_{\sun}$&$0.73\pm0.04$&$0.58\pm0.02$\\
          $Mv_1$&3.93&4.05\\
          $Mv_2$&5.17&5.42\\
          $R_1/R_{ZAMS}$&$1.37\pm0.01$&$1.69\pm0.01$\\
          $R_2/R_{TAMS}$&$1.34\pm0.01$&$1.31\pm0.01$\\
          $L_2/L_{TANS}$&$314.7$&$61.9$\\

            \hline
	\end{tabular}
	\caption{Light curve solution and absolute parameters for TYC 6995-813-1 and NSVS 13602901 }
	\end{table}
	
		 \begin{figure}[h!]
    \label{fig:RAAFIG1}
	\includegraphics[width=\columnwidth]{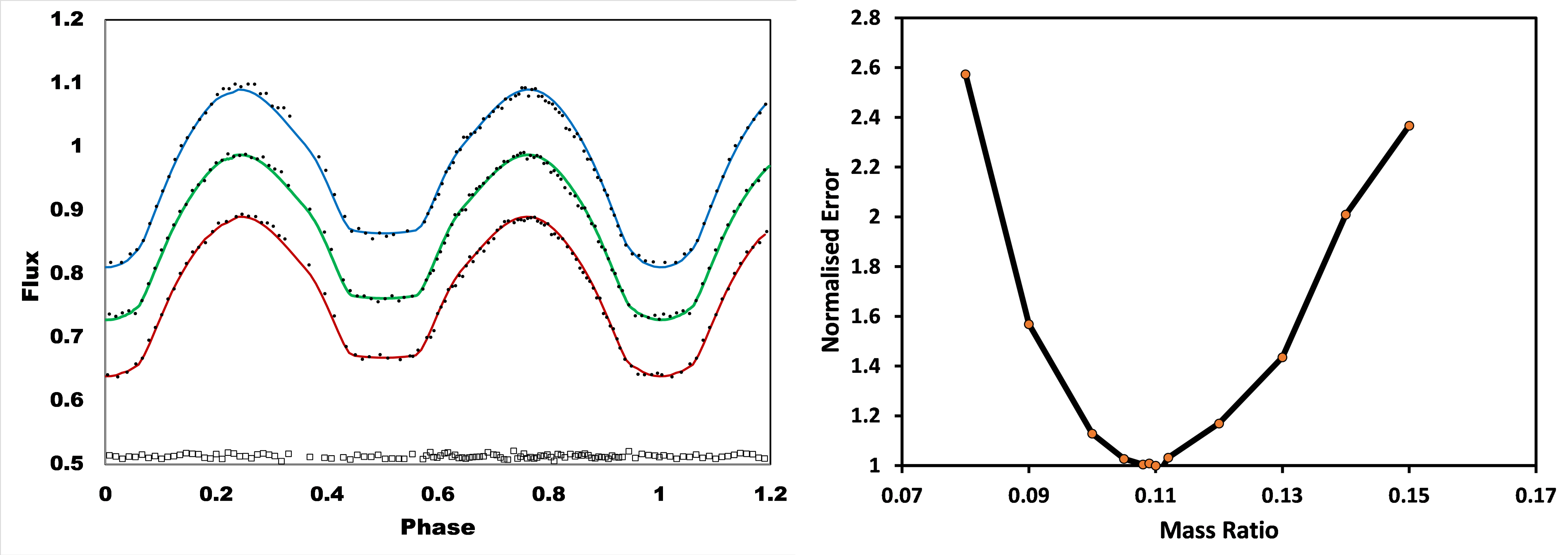}
    \caption{Observed and fitted light curves (left) and mass ratio search grid (right) for TYC 6995-813-1. For ease of display the blue curves are shifted up 0.1 units and the red curves shifted down 0.1 units. The check star flux is shown at the bottom of the light curve is shifted up by 0.1 units. The mass ratio search grid has been normalised to the minimum sum of the residuals ($2.216^{-2}$) as reported by the WD code }
    
\end{figure}

\begin{figure}[h!]
    \label{fig:RAAFIG2}
	\includegraphics[width=\columnwidth]{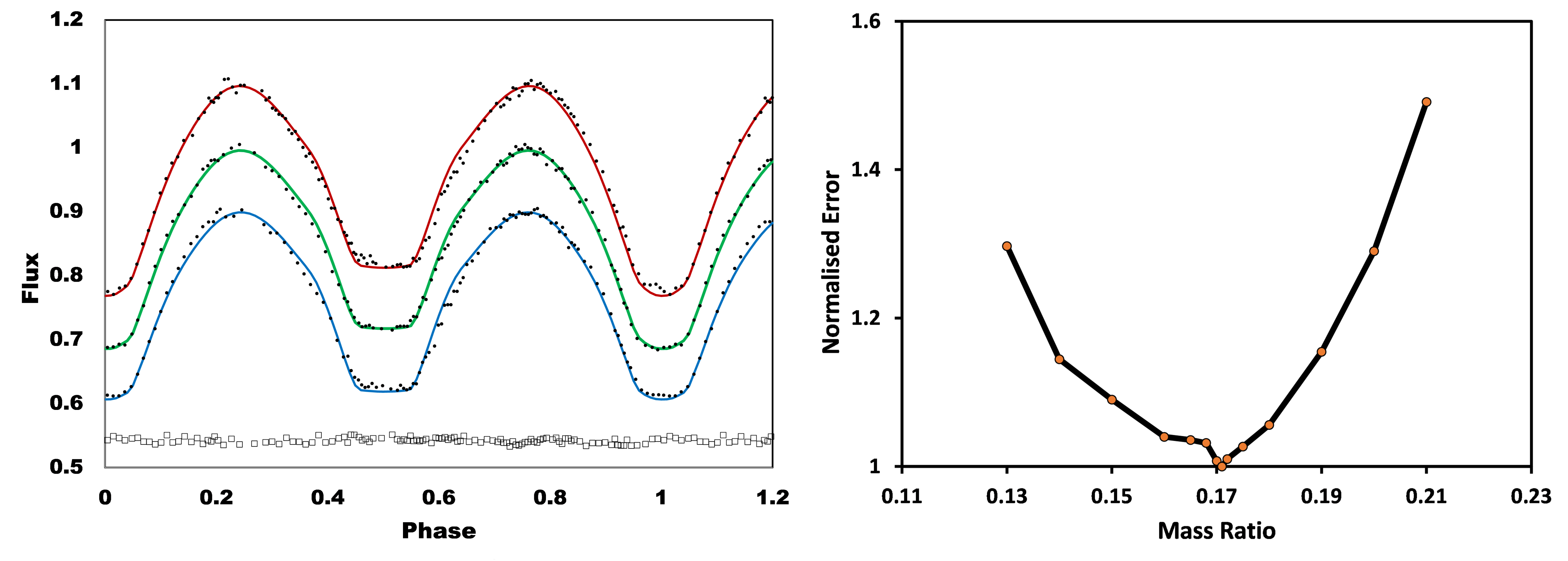}
    \caption{Observed and fitted light curves (left) and mass ratio search grid (left) for NSVS 13602901. For ease of display the blue curves are shifted up 0.1 units and the red curves shifted down 0.1 units. The check star flux is shown at the bottom of the light curve is shifted down by 0.05 units. The mass ratio search grid has been normalised to the minimum sum of the residuals ($3.51^{-1}$) as reported by the WD code}
    
\end{figure}

\begin{figure}[h!]
    \label{fig:RAAFIG3}
	\includegraphics[width=\columnwidth]{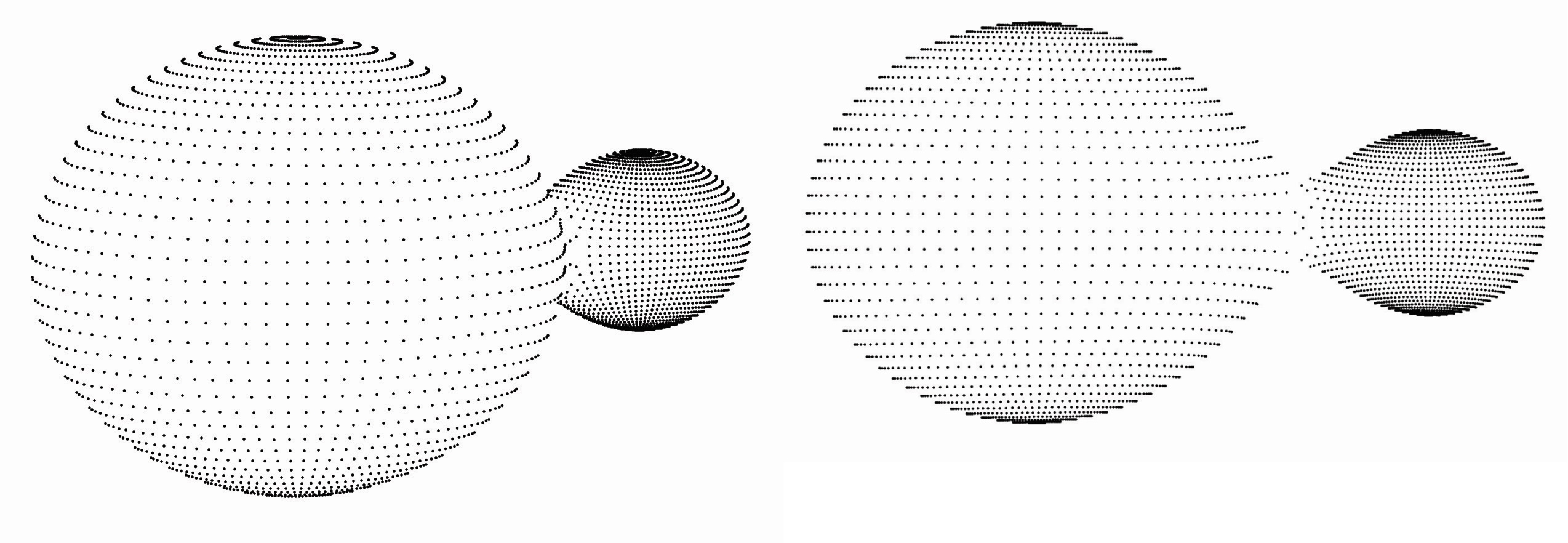}
    \caption{3D representations of TYC 6995-813-1 (left) and NSVS 13602901 (right) }
    
\end{figure}

\section{Merger Potential}
As noted above, contact binary systems are thought to merge when the orbital angular momentum falls below a critical (minimum) value. Recently, \citet{2021MNRAS.501..229W} showed that orbital instability is not independent of the masses of the components but critically dependent particularly on the mass of the primary. They concluded that instability is unlikely unless the separation between the components was less than a critical value ($A_{inst}$) and showed that this could be expressed a function of the mass ratio, radius of the primary and the gyration radii of the components and expressed as:

\begin{equation}
\label{eq:a-inst}
 \frac{A_{\mathrm{\scriptscriptstyle inst}}}{R_1} = \frac{q\frac{k_2^2}{k_1^2}{P}{Q} + \sqrt{(q\frac{k_2^2}{k_1^2}{P}{Q})^2 + 3 (1+q\frac{k_2^2}{k_1^2}{Q}^2) (q \frac{k_2^2}{k_1^2}{P}^2 + \frac{q}{(1+q)k_1^2})   }}{q \frac{k_2^2}{k_1^2}{P}^2 + \frac{q}{(1+q)k_1^2}}.
\end{equation}
Where $k_{1,2}$ is the gyration radius for the primary and secondary components and

\begin{equation}
P = \frac{0.49q^{2/3}-3.26667q^{-2/3}(0.27q -0.12q^{4/3})}{0.6q^{2/3} + \ln (1+ q^{1/3})} .
\end{equation}

\begin{equation}
Q = \frac{(0.27q -0.12q^{4/3})({0.6q^{-2/3} + \ln (1+ q^{-1/3})})}{0.15 (0.6q^{2/3} + \ln (1+ q^{1/3}))},
\end{equation}

Review of the modelled values of the gyration radii ($k$) of low mass ($0.4M_{\sun} < M < 1.4M_{\sun}$) rotating and tidally distorted ZAMS stars by \citet{2009A&A...494..209L} yields a simple linear relationship between the two as:
\begin{equation}
    k=-0.2455M + 0.5368
\end{equation}

We use this relationship to derive the values of $k_1$ (gyration radius of the primary) for the two systems. The secondary is a very low mass star and for a low mass secondary, the convective n=1.5 polytrope is an excellent approximation  \citep{2007MNRAS.377.1635A} for which $k_2=0.453$. This fixed value of $k_2$ and the variable value of $k_1$ from equation 8 and absolute parameters derived above were adopted to calculate the instability separation, using equation 5, for both systems reported. The instability separation of TYC 6995-813-1 is estimated as 2.35$R_{\sun}$ compared to the current separation only slightly higher at 2.46$R_{\sun}$ suggesting the the system is approaching instability. In contrast, the instability separation for NSVS 13602901 at 2.76$R_{\sun}$ is significantly smaller that the current separation of 3.05$R_{\sun}$ suggesting the system is likely to be stable.

\section{Summary and Conclusions}
First multi-band photometric analysis of two bright southern contact binary systems are presented. We find both systems show extreme low mass ratios with evidence of significant age given the relative brightness, size and density of the "current" secondary relative to main sequence counterparts of similar mass. As expected the primaries of both systems have absolute parameters not far removed from their ZAMS counterparts. Although neither system demonstrates a significant O'Connell effect at least one of the systems that was observed by the GALEX mission does have features suggestive of significant chromospheric activity. Post V1309 Sco, orbital instability of contact binary systems has received considerable attention and at least one of our systems, TYC 6995-813-1 would fall into the category of being a potential merger candidate. Among the most critical changes observed retrospectively for V1309 Sco was the exponential decline in its period in the years prior to the merger event \citep{2011A&A...528A.114T}. Given its relative brightness, TYC 6995-813-1, represents an easy target for long term follow up by smaller instruments such as campus telescopes or even advanced amateurs.

\normalem
\begin{acknowledgements}
This research has made use of the SIMBAD database, operated at CDS, Strasbourg, France.

\end{acknowledgements}

\bibliographystyle{raa}
\bibliography{bibtex}

\end{document}